\documentclass[12pt,a4paper]{article}

\usepackage[T1]{fontenc}
\usepackage{graphicx}
\usepackage{epsfig}
\usepackage{amsmath}
\usepackage{amssymb}
\usepackage{latexsym}

\usepackage[usenames]{color}

\begin{document}
\textwidth=135mm
 \textheight=200mm
\begin{center}
{\bfseries Kaon and pion femtoscopy at top RHIC energy in hydrokinetic model
\footnote{{\small Talk given at the the Sixth Workshop on Particle Correlations and Femtoscopy, BITP, Kiev, September 14 - 18,
2010.}}}
\vskip 5mm
Iu.A. Karpenko and Yu.M. Sinyukov
\vskip 5mm
{\small {\it Bogolyubov Institute for Theoretical Physics, Kiev, 03680, Ukraine}} \\
\end{center}
\vskip 5mm
\centerline{\bf Abstract}
The hydrokinetic model is applied to restore the initial conditions and
space-time picture of the matter evolution in central Au+Au
collisions at the top RHIC energy. The analysis is based on the
detailed reproduction of the pion and kaon momentum spectra and
femtoscopic data in whole interval of the transverse momenta studied
by both STAR and PHENIX collaborations. A good description of the pion and kaon
transverse momentum spectra and interferometry radii is reached with
both initial energy density profiles motivated by the Glauber
and Color Glass Condensate (CGC) models, however, at different energy densities.
\vskip 10mm
\section{Introduction}

In this letter we apply the HydroKinetic Model (HKM) proposed in \cite{PRL,PRC} and further developed in \cite{YadFizN, KarSin, KarSin2}
to an analysis of the femtoscopic measurements at RHIC for central
Au+Au collisions at the top energy $\sqrt{s}=200$ AGeV. Namely, we
analyze pion and kaon transverse momentum spectra and the $m_T$-
behavior of the pion and kaon interferometry radii to clarify, in
particularly, how these observables depend on the initial
conditions: Glauber and CGC-like.

Hydrokinetic approach combines the advantages
of the hydrodynamic approximation, where possible phase transitions
are encoded in the corresponding equation of state (EoS), and
microscopic approach, accounting for a non-equilibrated process of
the spectra formation due to gradual particle liberation. The
dynamical decoupling is described by the particle escape
probabilities in inhomogeneous hydrodynamically expanding systems in
the way consistent with the kinetic equations in the relaxation time
approximation for emission function. The basic hydrokinetic code,
proposed in \cite{PRC}, is modified now to include decays of
resonances into the expanding hadronic chemically non-equilibrated
system and, based on the resulting composition of the
hadron-resonance gas at each space-time point, to calculate the
equation of state (EoS) in a vicinity of this point. The obtained
local EoS allows one to determine the further evolution of the
considered fluid elements.


 \section {Model description and results}
Let us briefly describe the main features of the model.
Our results are all related to the central rapidity slice where we
use the boost-invariant Bjorken-like initial condition in longitudinal direction.

\textit{Initial conditions: }Two models of initial conditions for hydrodynamic expansion were used: Glauber model and CGC-like model.
In former one, initial energy density in the
transverse plane is proportional to the participant nucleon density\cite{Kolb}.
For CGC model, the procedure described in \cite{KarSin, sin1} give us the energy profile in the transverse plane:
 \begin{equation}
 \epsilon(x_T)=\epsilon_0\frac{\rho^{3/2}(0,x_T)}{\rho^{3/2}_0},
 \end{equation}
where the number of participants $\rho$ is defined in the same way as in Glauber model.
The parameter $\epsilon_0\equiv \epsilon(b=0,{\bf
x}_T=0)$ -- the maximal energy density at the initial moment of
thermalization -- is the first fitting parameter of HKM for both IC cases.

Following the ideas about the pre-thermal flow development \cite {sin1} we take a ``conventional'' proper-time of thermalization,
$\tau_i=$1 fm/c and non-zero transverse flow already developed at the thermalization time.
The initial transverse rapidity profile is supposed to be linear in
radius $r_T$:
\begin{equation}
y_T=\alpha\frac{r_T}{R_T},\quad \text{where}\quad R_T=\sqrt{<r_T^2>}
 \label{yT},
\end{equation}
here $\alpha$ is the second fitting parameter in our model. Note that the fitting
parameter $\alpha$ should include also a positive correction for
underestimated resulting transverse flow since in this work we did
not account in direct way for the viscosity effects \cite{Teaney}
neither at QGP stage nor at hadronic one. In formalism of HKM
\cite{PRC} the viscosity effects at hadronic stage are incorporated
in the mechanisms of the back reaction of particle emission on
hydrodynamic evolution which we ignore in current calculations.
Since the corrections to transverse flows which depend on unknown
viscosity coefficients are unknown, we use fitting parameter
$\alpha$ to describe the "additional unknown portions" of flows,
caused both factors: by a developing of the pre-thermal flows and
the viscosity effects in quark-gluon plasma.

\textit{Equation of state: }At high temperatures corresponding to the QGP phase and crossover transition
to hadron phase we use a realistic EoS \cite{Laine} adjusted to the
lattice QCD results for zero barionic chemical potential so that it
is matched with an ideal chemically equilibrated multicomponent
hadron resonance gas at $T_c=175$ MeV.
To take into account a conservation of the net baryon number, electric charge and
strangeness in the QGP phase, we make corrections to
thermodynamic quantities for nonzero chemical potentials, as proposed in \cite{karsch_nonzero}.

The chemical freeze-out temperature $T_{ch}=165$ MeV is used, with corresponding chemical
potentials $\mu_B$ =29 MeV, $\mu_S$ =7 MeV, $\mu_E$ =-1 MeV and also
the strangeness suppression factor $\gamma_S=0.935$ which are
dictated by 200A GeV RHIC particle number ratios analysis done in
the statistical model \cite{Becattini, PBM}.

At the chemical freeze-out temperature $T_{ch}$ the EoS is matched with ideal Boltzmann
hadronic resonance gas which includes $N=359$ hadron states made of
u, d, s quarks with masses up to 2.6 GeV (the same set of hadrons as used in \cite{Amelin}).
The EoS used at temperatures below chemical freeze-out is chemically non-equilibrated and takes into account the change of chemical composition due to decays of resonances in the form:
\begin{equation}
    \partial_\mu(n_i(x) u^\mu(x))=-\Gamma_i n_i(x) + \sum\limits_j b_{ij}\Gamma_j
    n_j(x)\label{decay}
\end{equation}
here one neglects a thermal motion of the resonance $j$, that can be
justified because  post (chemical) freeze-out temperatures are much
less than the mass of the lightest known resonance. Also, equations for the hydrodynamic evolution are written under
supposition of an instant thermalization of the decay products, that
is consistent with the ideal fluid  approximation (mean free path is
zero).

\textit{Spectra formation: }During the hadron phase, hadrons are permitted to leave the system. The process of particle emission is described by the means of emission function, which, for (quasi-)stable particles is  expressed through the gain term $G_i^\text{gain}(x,p)$ in Boltzmann equations and escape probabilities \cite{PRL,PRC}. In this formalism, particle emission is formed by the particles that undergo their last interaction or already free initially. Thus, emission depend on the collision rate of given particle with hadron medium, which is calculated in a standard way using the total hadron cross-sections taken in a way similar to that in UrQMD code.

\textit{Results and discussion: }The results of the HKM for the pion and kaon spectra, interferometry
radii and $R_{out}/R_{side}$ ratio are presented in Fig. 1. 
The best fit for the Glauber IC is reached at the following
values of the two fitting parameters related to the proper time
$\tau = 1$ fm/c: $\epsilon_0 = 16.5$ GeV/fm$^3$ ($\langle \epsilon
\rangle = 11.69$ GeV/fm$^3$) and parameter of the initial velocity
defined by (\ref{yT}), $\alpha$ = 0.248 ($\langle v_T \rangle =
0.224$). In the case of the CGC-like initial conditions $R_T = 3.88$
fm, the fitting parameters leading to the best data description are
$\epsilon_0 = 19.5$ GeV/fm$^3$ ($\langle \epsilon \rangle = 13.22$
GeV/fm$^3$) and $\alpha$ = 0.23 ($\langle v_T \rangle = 0.208$). Since
the temperature and baryonic chemical potential at chemical
freeze-out, which are taken from the analysis of the particle number
ratios \cite{Becattini}, are is more suitable for the STAR
experiment, the HKM results for kaon spectra are good for the STAR
data but not so much for the PHENIX ones. Note also that, in spite
of other studies (e.g., \cite{Broniowski}), we compare our results
for the interferometry radii within the whole measured interval of
$p_T$ covered at the top RHIC energy. Finally, one can conclude from
Fig. 1 that the description of pion and kaon spectra and space-time
scales is quite good for both IC, the Glauber and CGC. It is worth
noting, however, that the two fitting parameters $\alpha$ and
$\epsilon_0$ are various by 10-20$\%$ for different IC, as it is
described above.

The special attention acquires a good description of the pion and
kaon longitudinal radii altogether with $R_{out}/R_{side}$  ratio,
practically, within the experimental errors. Such an achievement
means that the HKM catches the main features of the matter evolution
in A+A collisions and correctly reproduces the homogeneity lengths
in the different parts of the system which are directly related to
the interferometry radii at the different momenta of the pairs
\cite{Sin}.

\begin{figure}
\hspace{0cm}
\includegraphics[scale=0.8]{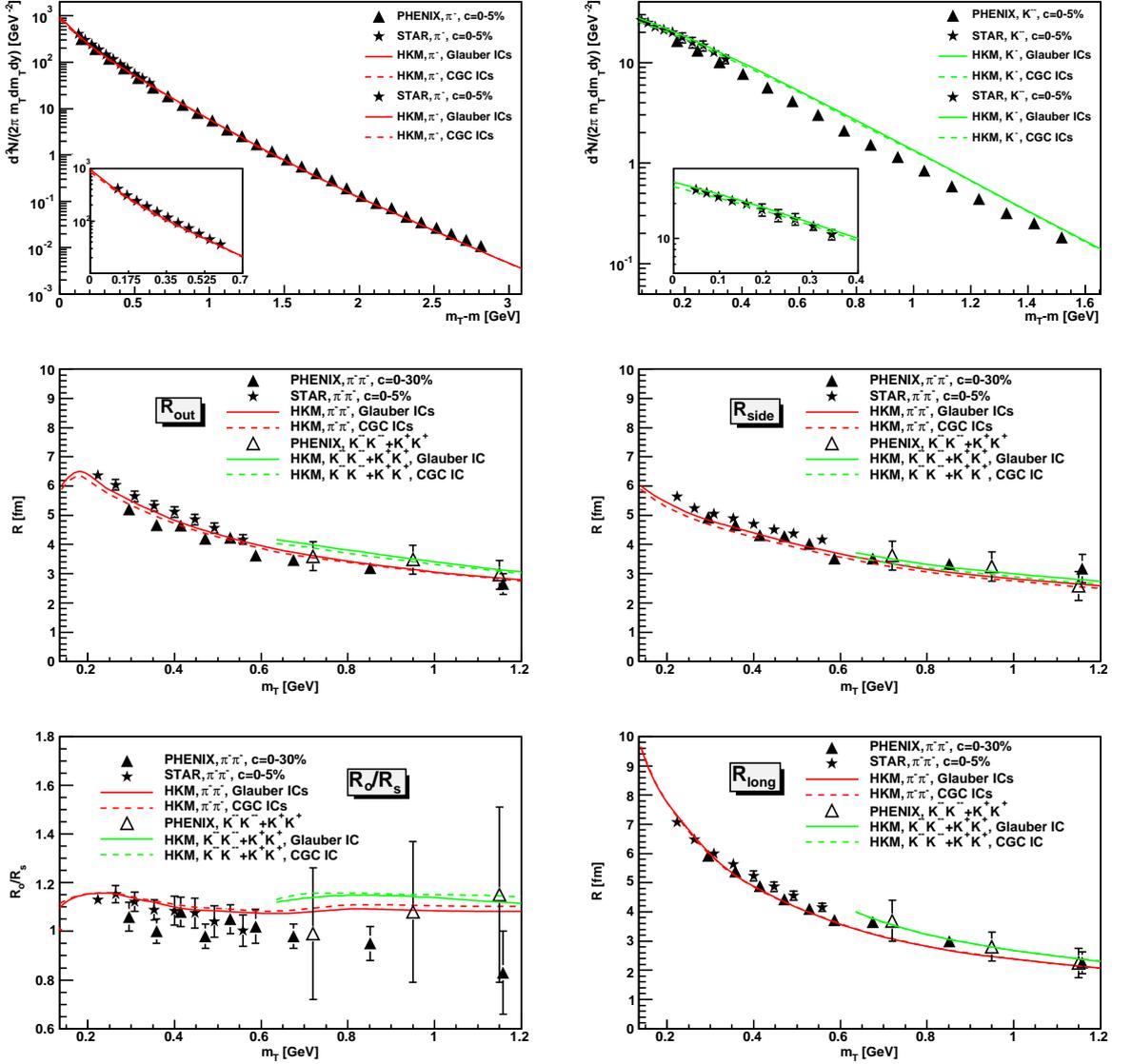}
\caption{The transverse momentum spectra of negative pions and
negative kaons, all calculated in the HKM model. The comparison only
with the STAR data are presented in the separate small plots. (Top).
The interferometry radii and $R_{out}/R_{side}$ ratio for
$\pi^{-}\pi^{-}$ pairs and mixture of $K^- K^-$ and $K^+ K^+$ pairs.
(Middle and bottom). The experimental data are taken from the STAR
\cite{star-spectra, star-hbt} and PHENIX \cite{phenix-spectra,
phenix-hbt, phenix-hbt-kaon} Collaborations.}
\end{figure}

\section{Conclusions}
In this letter, we show how hydro-kinetic model is applied to restore the initial conditions and space-time
picture of the matter evolution in central Au+Au collisions at the
top RHIC energy. The analysis, which is based on a detailed
reproduction of the pion and kaon momentum spectra and measured
femtoscopic scales, demonstrates that basically the pictures of the
matter evolution and particle emission are similar at both Glauber
and CGC initial conditions (IC) with, however, the different initial
maximal energy densities: it is about 20\% more for the CGC initial
conditions. The main factors, which allows one to describe well
simultaneously the spectra and femtoscopic scales are: a relatively
hard EoS (crossover transition and chemically non-equilibrium
composition of hadronic matter), pre-thermal transverse flows
developed to thermalization time, an account for an "additional
portion" of the transverse flows due to the shear viscosity effect
and fluctuation of initial conditions, a correct description of a
gradual decay of the non-equilibrium fluid at the late stage of
expansion.

\section*{Acknowledgments}

The authors thank S.V. Akkelin for fruitful discussions. The researches were carried
  out in part within the scope of the EUREA: European Ultra Relativistic Energies Agreement (European Research Group
  GDRE: Heavy ions at ultrarelativistic energies) and is supported by the State Fund for
Fundamental Researches of Ukraine (Agreement,2011)  and National Academy of Sciences of Ukraine (Agreement,2011).

\end{document}